\documentclass[11pt,oneside]{article}
\usepackage{graphicx,amsmath,amssymb}
\newtheorem{lemma}{Lemma}
\newtheorem{theorem}{Theorem}
\newtheorem{corollary}{Corollary}
\newtheorem{conjecture}{Conjecture}
\newcommand{\qed}{\hbox{\rule[-2pt]{3pt}{6pt}}}
\begin{document}
\begin{center}
{\bf\sc\Large
Duality in finite-dimensional spin glasses}

\vspace{.5cm}

{Hidetoshi Nishimori}

\vspace{.5cm}

{\small Department of Physics, Tokyo Institute of Technology, Tokyo, Japan}

\vskip 1truecm
\date{}

\end{center}

\vskip 1truecm
\begin{abstract}\noindent
We present an analysis leading to a conjecture on the exact location of the
multicritical point in the phase diagram of spin glasses in finite dimensions.
The conjecture, in satisfactory agreement with a number of numerical results,
was previously derived using an ansatz emerging from duality and the replica method.
In the present paper we carefully examine the ansatz and reduce it to
a hypothesis on analyticity of a function appearing in the duality relation.
Thus the problem is now clearer than before from a mathematical point of view:
The ansatz, somewhat arbitrarily introduced previously,
has now been shown to be closely related
to the analyticity of a well-defined function.
\end{abstract}

\section{Introduction}
To establish reliable analytical theories of spin glasses has been one
of the most challenging problems in statistical physics for years.
The problem was solved for the mean-field model \cite{SK,Parisi,Guerra,Talagrand}.
Much less is known analytically for finite-dimensional spin glasses,
for which approximate methods including numerical simulations and
phenomenological theories \cite{Young} have been the main tools of investigation
in addition to a limited set of rigorous results
and exact solutions \cite{HN2001,NS}.

Our main interest in the present contribution does not lie directly in
the issue of the properties of the spin glass phase.
We instead will concentrate ourselves on the precise (and possibly
exact) determination of the structure of phase diagram of
finite-dimensional spin glasses.
This problem is of
practical importance for numerical studies since exact
values of transition points greatly facilitate reliable estimates
of critical exponents in finite-size scaling.

More precisely, recent developments \cite{NN,MNN,TN,TSN,NO} to derive a conjecture
on the exact location of the multicritical point in the phase diagram
will be analyzed from a different view point.
The conjecture was derived using the replica method,
gauge symmetry and a duality relation.
In addition it was necessary to introduce an ansatz
to identify the location of a singularity of the free energy.
The resulting conjecture for the transition point (multicritical
point) is nevertheless in satisfactory agreement with a number of numerical results,
which renders a strong support to the validity of our prescription.
In the present paper we present a more systematic analysis leading
to the ansatz, thus reducing the problem to that of the analyticity
proof of a function related to duality.
The ansatz was introduced somewhat arbitrarily previously.
The discussions in the present paper make it clear that the ansatz
is closely related to the analyticity of a well-defined function,
which paves a path toward the formal proof of the conjecture.

\section{Duality relation for the replicated system}
For simplicity,
let us consider the $\pm J$ Ising model  on the square lattice.
It is possible to apply the same line of argument
as presented below to other systems (non-Ising models and/or
other lattices) as will be mentioned in the final section.
The Hamiltonian is
\begin{equation}
  H=-J\sum_{\langle ij\rangle} \tau_{ij} S_i S_j,
\end{equation}
where $\tau_{ij}=\pm 1$ is a quenched random variable
with asymmetric distribution
and $J>0$. Periodic boundary conditions are imposed.
We accept the replica method in this paper and do not strive
to rigorously justify the validity of taking the limit $n\to 0$
in the end, where $n$ is the number of replicas.

\begin{figure}[hb]
\begin{center}
\includegraphics[width=10mm]{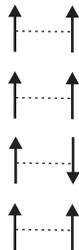}
\end{center}
\caption{The edge Boltzmann factor $x_k(K,K_p)$ represents the weight for
$k$ antiparallel spin pairs among $n$ nearest-neighboring pairs.
The case of $k=1, n=4$ is shown here.}
\label{fig:xk}
\end{figure}

The $n$-replicated partition function after configurational average
is a function of edge Boltzmann factors:
\begin{equation}
 [Z^n]\equiv Z_n(x_0(K,K_p), x_1(K,K_p),\cdots ,x_n(K,K_p)).
  \label{Zn}
\end{equation}
Here the square brackets denote the configurational average,
$K$ stands for $\beta J=J/k_BT$, and $K_p$ is a function of $p$
(the probability that $\tau_{ij}$ is 1) defined through $e^{-2K_p}=(1-p)/p$.
The $k$th edge Boltzmann factor $x_k(K,K_p)~(k=0,1,\cdots ,n)$
represents the configuration-averaged Boltzmann factor for
interacting spins with $k$ antiparallel spin pairs among $n$ nearest-neighbor
pairs for a bond (edge) as illustrated in Fig. \ref{fig:xk},
\begin{equation}
 x_k(K,K_p)=pe^{(n-2k)K}+(1-p)e^{-(n-2k)K}.
 \label{xk}
\end{equation}
%
The expression on the right-hand side of Eq. (\ref{Zn}) emphasizes the
fact that the system properties are uniquely determined by the
values of edge Boltzmann factors because the interactions (or equivalently, the
edge Boltzmann factors) do not depend on the bond index $\langle ij\rangle$
after the configurational average.

The formulation of duality transformation developed by Wu and Wang \cite{WuWang}
can be applied to the function $Z_n$ of Eq. (\ref{Zn}) to derive a dual
partition function on the dual square lattice with dual edge Boltzmann
factors. The result is, up to a trivial constant, \cite{MNN}
\begin{equation}
  Z_n(x_0,x_1,\cdots ,x_n)=Z_n(x_0^{*},x_1^{*},\cdots ,x_n^{*}).
  \label{Zn_duality}
\end{equation}
The dual Boltzmann factors are
discrete multiple Fourier transforms of the original Boltzmann factors
(which are simple combinations of plus and minus of the original Boltzmann
factors in the case of Ising spins):
\begin{equation}
  2^{n/2} x_m^{*} =  \sum_{k=0}^n  D_m^{k} \,   x_k, 
  \label{x_star}
\end{equation}
where the $D_m^{k}$ are the coefficients of the expansion
of  $(1-t)^m (1+t)^{n-m} $ :
\begin{equation}
 (1-t)^m (1+t)^{n-m} = 
\sum_{k=0}^n   D_m^{k} \, t^k,
 \label{t-expansion}
\end{equation}
or explicitly,
\begin{equation}
 D_m^k = \sum_{l=0}^k (-1)^l {m \choose l}{n-m \choose k-l}.
 \label{Dmk}
\end{equation}
For example, in the case of $n=2$, the dual Boltzmann factors are given as
\begin{equation}
 \begin{array}{rcl}
   2\, x_0^* &=&(x_0+x_1)+(x_1+x_2) =x_0+2x_1+x_2 \\
   2\, x_1^* &=&(x_0-x_1)+(x_1-x_2) =x_0-x_2 \\
   2\, x_2^* &=&(x_0-x_1)-(x_1-x_2) =x_0-2x_1+x_2 .
 \end{array}
 \label{dual_2}
\end{equation}

It will be useful to measure the energy from the all-parallel spin configuration
($k=0$) and, correspondingly, factor out the principal Boltzmann factors,
$x_0$ (left hand side) and $x_0^{*}$ (right hand side),
from both sides of Eq. (\ref{Zn_duality}).
Since the partition function is a homogeneous multinomial of edge Boltzmann
factors $(x_0,x_1,\cdots ,x_n)$ of order $N_B$ (the number of bonds),
the duality relation (\ref{Zn_duality}) can then be rewritten as
 \begin{equation}
  (x_0)^{N_B}\tilde{\mathcal{Z}}_n(u_1, u_2, \cdots , u_n)=(x_0^{*})^{N_B}
   \tilde{\mathcal{Z}}_n(u_1^{*}, u_2^{*}, \cdots , u_n^{*}),
   \label{Zn_duality2}
 \end{equation}
using the normalized edge Boltzmann factors $u_k$ and
$u_k^{*}~(k=1,2,\cdots ,n)$ defined by
$u_k=x_k/x_0$ and $u_k^{*}=x_k^{*}/x_0^{*}$.

The discussions so far have already been given in references \cite{MNN,TN,TSN}.
In those papers we went on to try to identify the multicritical point
by the fixed-point condition of the principal Boltzmann factor,
$x_0(K, K_p)=x_0^{*}(K,K_p)$, combined with the Nishimori line (NL)
condition $K=K_p$ \cite{HN2001}. 
The reason for the latter choice is that the multicritical point
is expected to lie on the NL \cite{DH}.
In this way an ansatz was made at this stage that the multicritical
point is given by the relation $x_0(K,K)=x_0^{*}(K,K)$.
The resulting expression for the location of the multicritical point
was confirmed to be exact in the cases of $n=1, 2$ and $n\to\infty$.
Extrapolation to the quenched limit $n\to 0$ gave results in
agreement with a number of independent numerical estimates
as listed in Table 1.
However, it was difficult to understand the mathematical origin
of the somewhat arbitrary-looking ansatz.

We develop an argument in the next section to justify the
above-mentioned relation $x_0(K,K)=x_0^{*}(K,K)$ as the fixed point
condition of duality relation for the replicated $\pm J$ Ising
model on the NL.

\section{Self-duality}

\begin{figure}[hb]
\begin{center}
\includegraphics[width=40mm]{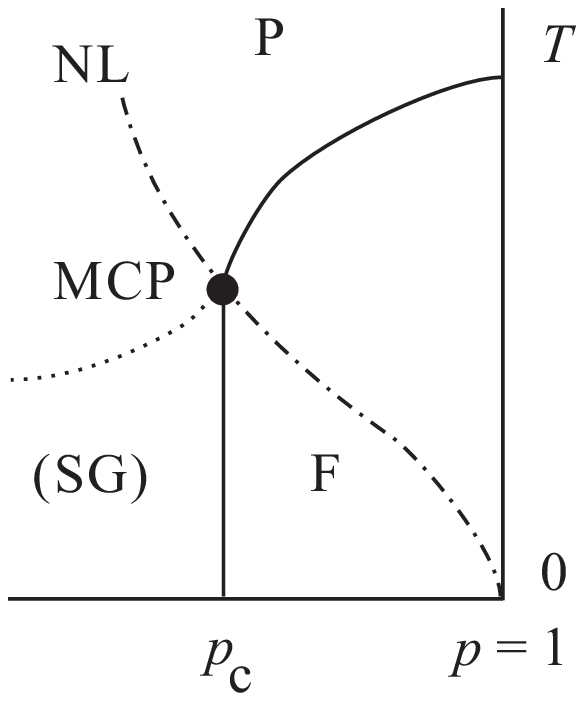}
\end{center}
\caption{A typical phase diagram of the $\pm J$ Ising model.
Shown dash-dotted is the NL.  Note that
the spin glass phase is believed not to exist in two dimensions.
The multicritical point is denoted as MCP.}
\label{fig:phase_diagram}
\end{figure}

The duality relation (\ref{Zn_duality2}) applies to an arbitrary
set of parameter values $(K,K_p)$ or an arbitrary point in the
phase diagram (Fig. \ref{fig:phase_diagram}).
However, we restrict ourselves to the NL ($K=K_p$) to investigate
the location of the multicritical point.
Then Eq. (\ref{Zn_duality2}) is expressed as
\begin{equation}
  (x_0(K))^{N_B}\tilde{\mathcal{Z}}_n(u_1(K),u_2(K),\cdots ,u_n(K))
 =(x_0^{*}(K))^{N_B}\tilde{\mathcal{Z}}_n(u_1^{*}(K),u_2^{*}(K),\cdots ,u_n^{*}(K))
 \label{Zn_duality3}
\end{equation}
since $x_0, u_1,\cdots, u_n$ and $x_0^{*}, u_1^{*},\cdots ,u_n^{*}$ are now
functions of a single variable $K$.

\begin{figure}[hb]
\begin{center}
\includegraphics[width=40mm]{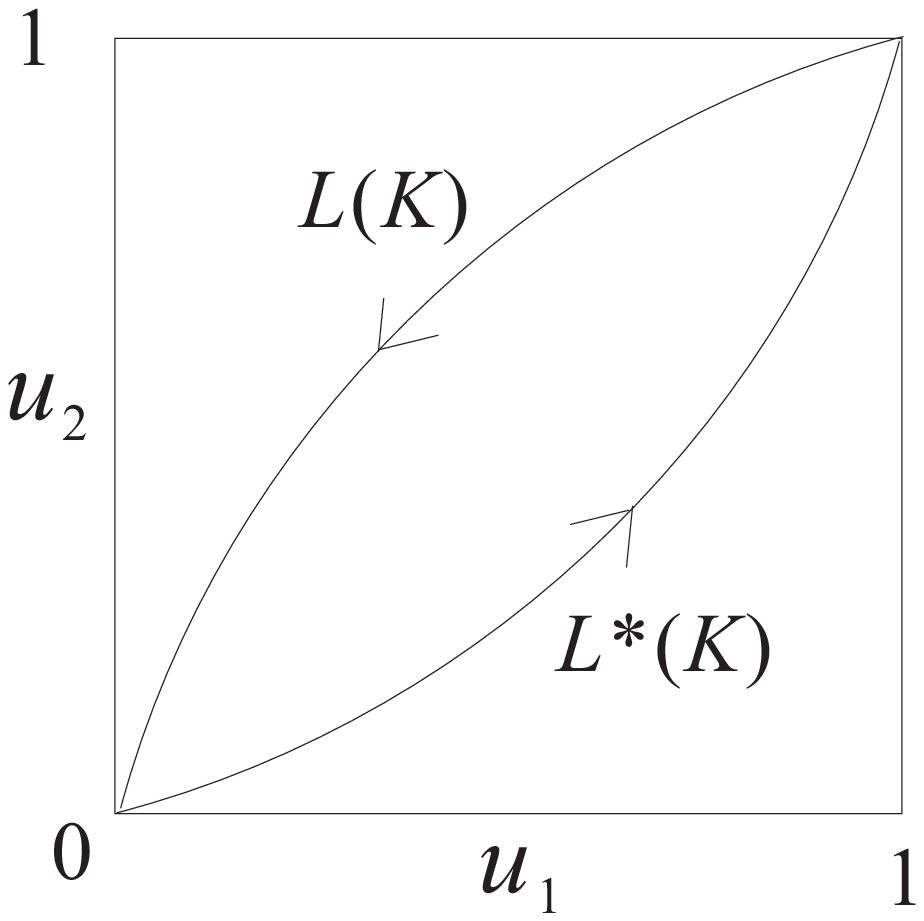}
\end{center}
\caption{Trajectories of the points $\mathcal{L}(K)=(u_1(K), u_2(K),\cdots ,u_n(K))$
and $\mathcal{L}^{*}(K)=(u_1^{*}(K), u_2^{*}(K),\cdots ,u_n^{*}(K))$,
projected onto the $(u_1,u_2)$ plane.  The arrows indicate the sense of motion
of $\mathcal{L}(K)$ and $\mathcal{L}^{*}(K)$ as $K$ changes from 0 to $\infty$.}
\label{fig:trajectory}
\end{figure}

It is difficult to apply directly the usual duality argument
(identification of a fixed point with the transition point assuming uniqueness of
the latter) to
Eq. (\ref{Zn_duality3}).
The reason is that $\tilde{\mathcal{Z}}_n$ is a multi-variable function:
The two trajectories representing $\mathcal{L}(K)\stackrel{\rm def}{=} 
(u_1(K), u_2(K),\cdots ,u_n(K))$
and  $\mathcal{L}^{*}(K)\stackrel{\rm def}{=}(u_1^{*}(K), u_2^{*}(K),\cdots ,u_n^{*}(K))$
do not in general coincide as depicted in Fig. \ref{fig:trajectory}.
In other words, there is no fixed point in the conventional
sense for the present system.

In spite of this difficulty of the apparent absence of a
fixed point, we now show that
it is still possible to devise a map which renders the present
replicated system self-dual with a fixed point.
To derive such a result, it is useful to note a few properties of the
quantities appearing in Eq. (\ref{Zn_duality3}).
\begin{lemma}
The normalized Boltzmann factor $u_k(K)$ is a monotone decreasing function of $K$
from $u_k(0)=1$ to $\lim_{K\to\infty}u_k(K)=0$.
The dual $u_k^{*}(K)$ is a monotone increasing function from
$u_k^{*}(0)=0$ to $\lim_{K\to\infty}u_k^{*}(K)=1$.
Here $k\in \{1,2,\cdots n\}$ and $n\in \mathbb{N}$.
\label{lemma1}
\end{lemma}
{\em Proof.}
From the definition (\ref{xk}) of $x_k$ and the NL condition $K=K_p$
(i.e. $e^{-2K}=(1-p)/p)$,
it is straightforward to verify that
\begin{equation}
  u_k(K)=\frac{e^{(n+1-2k)K}+e^{-(n+1-2k)K}}{e^{(n+1)K}+e^{-(n+1)K}},
  \label{uk}
\end{equation}
from which the result for $u_k(K)$ follows.
The dual are shown to satisfy \cite{MNN}
\begin{equation}
  u_k^{*}(K)=\left\{
   \begin{array}{ll}
    (\tanh K)^{k+1}& (k~{\rm odd})\\
    (\tanh K)^{k}& (k~{\rm even})
   \end{array}
  \right. ,
  \label{uks}
\end{equation}
which leads to the statement on $u_k^{*}$.
\hfil\qed

\begin{lemma}
The normalized partition function $\tilde{\mathcal{Z}}_n(u_1,\cdots ,u_n)$
is a monotone increasing continuous function of all of its arguments $u_1, \cdots ,u_n$
with the limiting values in the hypercube $[0,1]^n$
$\tilde{\mathcal{Z}}_n(0,\cdots ,0)=2$ and
$\tilde{\mathcal{Z}}_n(1,\cdots ,1)=2^{nN}$, where $N$ is the number of sites.
\label{lemma2}
\end{lemma}
{\em Proof.}
Since the normalized partition function is a multinomial of normalized
edge Boltzmann factors $u_1,\cdots ,u_n$ with positive
coefficients, the first half of the above statement is trivial.
To check the second half, we note that $u_1=\cdots =u_n=0$ corresponds
to the case where no antiparallel spin pairs are allowed in any replica at any
bond.  The only allowed spin configuration is the all-parallel (i.e. perfectly
ferromagnetic) one, for which we have set the energy 0 (or the edge Boltzmann
factor $u_0=1$) since the energy is measured from such a state
by dividing $x_k$ by $x_0$.
Taking into account the global inversion degeneracy, we conclude
that $\tilde{\mathcal{Z}}_n(0,\cdots ,0)$ is equal to $2u_0=2$.

Similarly, when $u_1=\cdots =u_n=1$ (corresponding to the high-temperature limit),
all spin configurations show up with equal probability.
Therefore the normalized partition function just counts the number of
possible spin configurations, yielding $2^{nN}$.
\hfil\qed

\vspace{2mm}

Now we are ready to step toward the main theorems.
\begin{theorem}
There exists a monotone decreasing function $d(K)$ with $\lim_{K\to 0}d(K)\to\infty$
and $lim_{K\to\infty}d(K)=0$ by which the value
of $\tilde{\mathcal{Z}}$ at $(u_1^{*}(K),\cdots, u_n^{*}(K))$
becomes equal to the value at $(u_1(d(K)),\cdots, u_n(d(K)))$,
\begin{equation}
\tilde{\mathcal{Z}}_n(u_1^{*}(K),\cdots ,u_n^{*}(K))=
\tilde{\mathcal{Z}}_n(u_1(d(K)),\cdots ,u_n(d(K))).
\end{equation}
\end{theorem}
{\em Proof.}
According to Lemma \ref{lemma1}, the curve $\mathcal{L}(K)=(u_1(K),\cdots, u_n(K))$
starts from the point $(1,\cdots ,1)$ and ends at
$(0,\cdots ,0)$ as $K$ increases from 0 to $\infty$.
Similarly, the curve $\mathcal{L}^{*}(K)=(u_1^{*}(K),\cdots, u_n^{*}(K))$ starts from
$(0,\cdots ,0)$ and ends at $(1,\cdots ,1)$.
Thus $\tilde{\mathcal{Z}}_n(u_1(K),\cdots ,u_n(K))\stackrel{\rm def}{=} \mathcal{U}(K)$
is continuous and monotone decreasing from $\mathcal{U}(0)=2^{nN}$ to
$\lim_{K\to\infty}\mathcal{U}(K)=2$ by Lemma \ref{lemma2}.
Similarly,
$\tilde{\mathcal{Z}}_n(u_1^{*}(K),\cdots ,u_n^{*}(K))\stackrel{\rm def}{=} \mathcal{V}(K)$
is continuous and monotone increasing from $\mathcal{V}(0)=2$ to
$\lim_{K\to\infty}\mathcal{V}(K)=2^{nN}$.
Consequently there exists a monotone decreasing function $d(K)$
(satisfying $\lim_{K\to 0}d(K)\to\infty$ and $lim_{K\to\infty}d(K)=0$)
that relates the values of $\mathcal{U}$ and $\mathcal{V}$ such that
$\mathcal{U}(d(K))=\mathcal{V}(K)$.
\hfil\qed

\begin{corollary}
The normalized partition function satisfies the duality relation
\begin{equation}
  (x_0(K))^{N_B}\tilde{\mathcal{Z}}_n(u_1(K),\cdots ,u_n(K))
  =  (x_0^{*}(K))^{N_B}\tilde{\mathcal{Z}}_n(u_1(d(K)),\cdots ,u_n(d(K))),
\label{Zn_duality4}
\end{equation}
where $d(K)$ is the monotone decreasing function shown to exist in Theorem 1.
\end{corollary}
{\em Proof.}
Immediate from Eq. (\ref{Zn_duality3}) and Theorem 1.
\begin{theorem}
The normalized partition function is self-dual with a fixed point $K_0$ given by
$K_0=d(K_0)$, which is equivalent to $x_0(K_0)=x_0^{*}(K_0)$.
\end{theorem}
{\em Proof.}
The first half of the statement is immediate from Corollary 1.
The second half comes from the observation that the values of $\tilde{\mathcal{Z}}_n$
on both sides of Eq. (\ref{Zn_duality4}) become identical at the fixed point
$K_0=d(K_0)$, thus making the prefactors, $x_0(K)$ and $x_0^{*}(K)$,
equal to each other.
\begin{corollary}
Assume that $d(K)$ has no singularity for $K\in[0,\infty)$.
If the free energy per site
of the replicated system has a unique singularity at some $K=K_c$
in the thermodynamic limit, then $K_c$ is equal to $K_0$ of Theorem 2.
\end{corollary}
{\em Note.}
As long as $d(K)$ is analytic, the above statement is the same one
as the usual duality argument for the ferromagnetic Ising model on the square
lattice, in which $d(K)$ is $-(1/2)\log \tanh K$.
However, if $d(K)$ happens to be singular at $K_1$ for example,
this singularity would be reflected in the singularity
of the free energy (away from the fixed point $K_0$)
through the relation (\ref{Zn_duality4}):
The free energy per spin derived from the right-hand side will be
singular at $K_1$ reflecting the singularity of $d(K)$ there.
This causes a singularity of the free energy derived from the left-hand
side at $K_1$.

We have been unable to prove analyticity of $d(K)$.
This is one of the reasons that the following two statements are conjectures.
The other reasons include the validity of the replica method and the
absence of a formal proof for the existence of the multicritical point on the NL.
\begin{conjecture}{\rm \cite{MNN,NN}}
The exact location of the multicritical point for the $n$-replicated $\pm J$ Ising model
on the square lattice is given by the relation $x_0=x_0^{*}$ with $K=K_p$, i.e.
\begin{equation}
 e^{(n+1)K_c}+e^{-(n+1)K_c}=2^{-n/2}(e^{K_c}+e^{-K_c})^n.
 \label{MCPn}
\end{equation}
\end{conjecture}
{\em Note.}
The explicit expressions of $x_0$ and $x_0^{*}$ are given in
Eqs. (\ref{xk}) and (\ref{x_star}). See also references \cite{MNN,TN,TSN}.
\begin{conjecture}{\rm \cite{MNN,NN}}
The exact location of the multicritical point for the $\pm J$ Ising model
on the square lattice with quenched randomness is given by the formula
\begin{equation}
 -p_c\log p_c-(1-p_c)\log (1-p_c)=\frac{\log 2}{2}.
\end{equation}
\end{conjecture}
{\em Proof.}
The limit $n\to 0$ of Eq. (\ref{MCPn}) and the NL condition
$K=K_P$ yields the above formula.
%

\section{Concluding remarks}

The main physical conclusions, Conjecture 1 and Conjecture 2, are not new.
The significance of the present paper is that we have {\em derived}
the ansatz to identify the multicritical point, $x_0(K)=x_0^{*}(K)$,
from the assumption of analyticity of the function $d(K)$,
thus hopefully coming a little closer to the formal proof of the conjecture.

In this paper we have limited ourselves to the $\pm J$ Ising model
on the square lattice for simplicity.
It is straightforward to apply the same type of argument to other lattices
and other models.
For example, models on the square lattice (such as the Gaussian Ising spin glass
and the random chiral Potts model) can be treated very similarly: The differences
lie only in the explicit expressions of $x_k$ as given in section 2.9 of
reference \cite{MNN} and section 4.1 of reference \cite{TSN}.
Also, the duality structure of the four-dimensional random plaquette gauge model
is exactly the same as the $\pm J$ Ising model on the square lattice \cite{TN},
and therefore the present analysis applies without change.
In the case of mutually dual pairs of lattices, such as the triangular and
hexagonal lattices or the three-dimensional $\pm J$ Ising model and the
three-dimensional random-plaquette gauge model, a simple generalization suffices
that refers to the composite duality relation of the two systems,
$x_0(K_1)x_0(K_2)=x_0^{*}(K_2)x_0^{*}(K_1)$, where $K_1$ is the critical point for
one of the systems
and $K_2$ is for its dual, as detailed around Eq. (17) of reference \cite{TSN}.

Let us next give a few remarks on the comparison with numerics in Table 1.
As is seen there, our conjecture agrees with a number of numerical results
but lies slightly outside error bars
for some instances for the square lattice,
0.8907(2) \cite{MC} and 0.8906(2) \cite{HPP} vs.
0.889972 of our conjecture.
A similar situation is observed for three pairs of mutually dual hierarchical lattices
analyzed in \cite{HB}: The sum of the values of binary entropy
$H(p)=-p\log_2 p-(1-p)\log_2 (1-p)$ for the pair of mutually dual multicritical points
is exactly equal to 1 according to our conjecture, whereas the numerical results
are not precisely unity, 1.0172, 0.9829 and 0.9911.
We have no definite ideas at the present moment where these subtle differences come from.
Further investigations are necessary.

The final remark is on the transition points away from the NL (or the shape
of the phase boundary away from the multicritical point).
We expect (but cannot prove) that the relation $x_0(K,K_p)=x_0^{*}(K,K_p)$
gives the true critical point only on the NL.
An important reason is that the limiting behavior of the transition point
as $p\to 1$,
derived from the ansatz $x_0(K,K_p)=x_0^{*}(K,K_p)$, shows a deviation from
a perturbational result, see \cite{NN}.
The particularly high symmetry of the system on the NL \cite{GAL}
could be a reason for the success only on the NL.

\begin{table}[hb]
\begin{center}
{ \begin{tabular}{@{}llclc@{}}
\hline
  Model & Numerical estimates &Reference  &  Our conjecture & Reference\\
\hline
  SQ Ising                  & 0.8900(5) & \cite{dQ}         & 0.889972&\cite{NN,MNN} \\
                            & 0.8894(9) & \cite{IO}         &  & \\
                            & 0.8907(2) & \cite{MC}         &  & \\
                            & 0.8906(2) & \cite{HPP}      &  & \\
                            & 0.8905(5) & \cite{AR}         &  & \\
  SQ Gaussian               & 1.00(2)&\cite{OzekiG}        & 1.021770&\cite{NN,MNN} \\
  SQ 3-Potts                & 0.079-0.080 &\cite{JP}         & 0.079731&\cite{NN,MNN} \\
  $4d$ gauge (RPGM)         & 0.890(2)&\cite{Ichinose}       & 0.889972&\cite{TN} \\
  TR                        & 0.8355(5)&\cite{dQ}          & 0.835806&\cite{NO} \\
  HEX                       & 0.9325(5)&\cite{dQ}          & 0.932704&\cite{NO} \\
  $3d$ Ising (RBIM)         & 0.7673(3)$(=p_{c1})$&\cite{OI}          & ---& \\
  $3d$ gauge (RPGM)         & 0.967(4)$(=p_{c2})$&\cite{OAIM-WHP}     & --- &\\
  \quad RBIM+RPGM                 & $H(p_{c1})+H(p_{c2})=0.99(2)$& & $H(p_{c1})+H(p_{c2})=1$
                            &\cite{TSN}\\
  Hierarchical 1 (H1)       & 0.8265$(=p_{c3})$&\cite{HB}          & --- &\\
  Dual of H1 (dH1)          & 0.93380$(=p_{c4})$&\cite{HB}     & --- &\\
  \quad H1+dH1                    & $H(p_{c3})+H(p_{c4})=1.0172$&&$H(p_{c3})+H(p_{c4})=1$
                            &\cite{TSN}\\
  Hierarchical 2 (H2)       & 0.8149$(=p_{c5})$&\cite{HB}          & --- &\\
  Dual of H2 (dH2)          & 0.94872$(=p_{c6})$&\cite{HB}     & --- &\\
  \quad H2+dH2                    & $H(p_{c5})+H(p_{c6})=0.9829$& & $H(p_{c5})+H(p_{c6})=1$&
                                   \cite{TSN}\\
  Hierarchical 3 (H3)       & 0.7527$(=p_{c7})$&\cite{HB}          & --- &\\
  Dual of H3 (dH3)          & 0.97204$(=p_{c8})$&\cite{HB}     & ---& \\
  \quad H3+dH3                    & $H(p_{c7})+H(p_{c8})=0.9911$ && $H(p_{c7})+H(p_{c8})=1$&
             \cite{TSN}\\
  Hierarchical 4            & 0.8902(4) &\cite{Nobre}          & 0.889972 & \cite{MNN,TN}\\

\hline
 \end{tabular}
}
\caption{Location of the multicritical point by recent
numerical studies and our conjecture.
SQ stands for the square lattice, and TR/HEX for the triangular/hexagonal lattices,
respectively.
RBIM stands for the random-bond Ising model and RPGM is
for the random-plaquette gauge model.
See reference \cite{HB} for the explicit definition of three types of hierarchical
lattices (H1, H2, H3) and their duals.
Hierarchical 4 is a self-dual lattice, for which the analysis in \cite{MNN,TN}
applies directly (though not stated explicitly in these references).
The values are for $p_c$ of the $\pm J$ model
except for the Gaussian randomness on the square lattice
for which the values 1.00(2) and 1.021770 are for $J_{0c}/J$.
Spin variables are Ising excepting the three-state Potts model as indicated.
The symbol $H(p)$ is for the binary entropy $H(p)=-p\log_2 p-(1-p)\log_2 (1-p)$.}
\end{center}
\end{table}



\begin{thebibliography}{CMN}

\bibitem{SK}
D. Sherrington and S. Kirkpatrick, {\it Phys. Rev. Lett.} {\bf 35},
 1792 (1975).

\bibitem{Parisi}
M. M\'ezard, G. Parisi and M. A. Virasoro, {\it Spin Glass Theory and Beyond}
(World Scientific, Singapore, 1987).

\bibitem{Guerra}
F. Guerra, {\it Commun. Math. Phys.} {\bf 233}, 1 (2003).

\bibitem{Talagrand}
M. Talagrand, (this issue).

\bibitem{Young}
A. P. Young (ed), {\it Spin Glasses and Random Fields}
(World Scientific, Singapore, 1997).


\bibitem{HN2001} H. Nishimori, {\it Statistical Physics of Spin Glasses
and Information Processing: An Introduction}
(Oxford, Oxford, 2001).

\bibitem{NS}
C. W. Newman and D. L. Stein, (this issue).

\bibitem{NN} H. Nishimori and K. Nemoto, {\it J. Phys. Soc. Jpn.}
 {\bf 71}, 1198 (2002).
 
\bibitem{MNN} J.-M. Maillard, K. Nemoto and H. Nishimori {\it J. Phys.}
 {\bf A36}, 9799 (2003).

\bibitem{TN} K. Takeda and H. Nishimori, {\it Nucl. Phys.} {\bf B686},
 377 (2004).

\bibitem{TSN} K. Takeda, T. Sasamoto and  H. Nishimori, {\it J. Phys.}
 {\bf A38}, 3751 (2005).

\bibitem{NO} H. Nishimori and M. Ohzeki, {\it J. Phys. Soc. Jpn.}
 {\bf 75}, 034004 (2006).
 
\bibitem{WuWang}F. Y. Wu and Y. K. Wang,
{\it J. Math. Phys.} {\bf 17}, 439 (1976).

\bibitem{DH} P. Le Doussal and A. B. Harris, {\it Phys. Rev.} {\bf B40}, 9249 (1989).

\bibitem{dQ} S. L. A. de Queiroz, {\it Phys. Rev.} {\bf B73}, 064410 (2006).


\bibitem{IO} N. Ito and Y. Ozeki, {\it Physica} {\bf A321}, 262 (2003).

\bibitem{MC} F. Merz and J. T. Chalker, {\it Phys. Rev.} {\bf B65}, 054425 (2002).

\bibitem{HPP} A. Honecker, M. Picco and P. Pujol, {\it Phys. Rev. Lett.}
 {\bf 87}, 047201 (2001).

\bibitem{AR} F. D. A. Aar{\~a}o Reis, S. L. A. de Queiroz and R. R. dos Santos,
 {\it Phys. Rev.} {\bf B60}, 6740 (1999).

\bibitem{OzekiG} Y. Ozeki, (private communication)

\bibitem{JP} J. L. Jacobsen and M. Picco, {\it Phys. Rev.} {\bf E65}, 026113 (2002).

\bibitem{Ichinose} G. Arakawa, I. Ichinose, T. Matsui and K. Takeda,
{\it Nucl. Phys.} {\bf B709}, 296 (2005).

\bibitem{OI} Y. Ozeki and N. Ito, {\it J. Phys.} {\bf A31}, 5451 (1998).

\bibitem{OAIM-WHP} T. Ohno, G. Arakawa, I. Ichinose and T. Matsui,
   {\it Nucl. Phys.} {\bf B697}, 462 (2004); C. Wang, J. Harrington and J. Preskill,
   {\it Ann. Phys.} {\bf 303}, 31 (2003).

\bibitem{HB} M. Hinczewski and A. N. Berker, {\it Phys. Rev.} {\bf B72}, 144402 (2005).

\bibitem{Nobre} F. D. Nobre, {\it Phys. Rev.} {\bf E64}, 046108 (2001).

\bibitem{GAL}I. A. Gruzberg, N. Read and A. W. W. Ludwig, {\it Phys. Rev.}
{\bf 63}, 104422 (2001).



\end{thebibliography}
\end{document}